\documentclass[10pt,conference]{IEEEtran}

\usepackage{enumitem}
\usepackage{url}
\usepackage{cite}
\usepackage{amsmath,amssymb,amsfonts}
\usepackage{algorithmic}
\usepackage{graphicx}
\usepackage{textcomp}
\usepackage{xcolor}
\usepackage{listings}
\usepackage{subcaption}
\usepackage{colortbl}
\usepackage{booktabs}
\usepackage{multirow}
\usepackage{arydshln}
\usepackage[most]{tcolorbox}
\def\BibTeX{{\rm B\kern-.05em{\sc i\kern-.025em b}\kern-.08em
    T\kern-.1667em\lower.7ex\hbox{E}\kern-.125emX}}
    \makeatletter
\newcommand{\linebreakand}{%
  \end{@IEEEauthorhalign}
  \hfill\mbox{}\par
  \mbox{}\hfill\begin{@IEEEauthorhalign}
}
\makeatother

\newcommand{\circnum}[1]{\textcircled{\tiny#1}}

\definecolor{dkgreen}{rgb}{0,0.6,0}
\definecolor{gray}{rgb}{0.5,0.5,0.5}
\definecolor{mauve}{rgb}{0.58,0,0.82}

\definecolor{mygray}{gray}{1}

\begin{document}

\title{Improving Dynamic Specification Inference with LLM-Generated Counterexamples}

\author{
    \IEEEauthorblockN{Agust\'in Balestra\IEEEauthorrefmark{1}, 
                      Agust\'in Nolasco\IEEEauthorrefmark{1}\IEEEauthorrefmark{5}, 
                      Facundo Molina\IEEEauthorrefmark{2}, \\
                      Diego Garbervetsky\IEEEauthorrefmark{3}\IEEEauthorrefmark{5}\IEEEauthorrefmark{6},
                      Renzo Degiovanni\IEEEauthorrefmark{4}, and 
                      Nazareno Aguirre\IEEEauthorrefmark{1}\IEEEauthorrefmark{5}\IEEEauthorrefmark{6}}
    
    \IEEEauthorblockA{\IEEEauthorrefmark{1}University of Rio Cuarto, Rio Cuarto, Argentina, \textsf{$\lbrace$ebalestra, nolasco, naguirre$\rbrace$@dc.exa.unrc.edu.ar}}
    \IEEEauthorblockA{\IEEEauthorrefmark{2}Complutense University of Madrid, Madrid, Spain, \textsf{facundom@ucm.es}}
    \IEEEauthorblockA{\IEEEauthorrefmark{3}University of Buenos Aires, Buenos Aires, Argentina, \textsf{diegog@dc.uba.ar}}
    \IEEEauthorblockA{\IEEEauthorrefmark{4}Luxembourg Institute of Science and Technology, Luxembourg, \textsf{renzo.degiovanni@list.lu}}
    \IEEEauthorblockA{\IEEEauthorrefmark{5}National Council for Scientific and Technical Research (CONICET), Argentina}
    \IEEEauthorblockA{\IEEEauthorrefmark{6}Guangdong Technion-Israel Institute of Technology, Shantou, China}   
    }

\maketitle

\begin{abstract}
Contract assertions, such as preconditions, postconditions, and invariants, play a crucial role in software development, enabling applications such as program verification, test generation, and debugging. Despite their benefits, the adoption of contract assertions is limited, due to the difficulty of manually producing such assertions. Dynamic analysis-based approaches, such as Daikon, can aid in this task by inferring expressive assertions from execution traces. However, a fundamental weakness of these methods is their reliance on the thoroughness of the test suites used for dynamic analysis. When these test suites do not contain sufficiently diverse tests, the inferred assertions are often not generalizable, leading to a high rate of invalid candidates (false positives) that must be manually filtered out. 

In this paper, we explore the use of large language models (LLMs) to automatically generate tests that attempt to invalidate generated assertions. Our results show that state-of-the-art LLMs can generate effective counterexamples that help to discard up to 11.68\% of invalid assertions inferred by SpecFuzzer. Moreover, when incorporating these LLM-generated counterexamples into the dynamic analysis process, we observe an improvement of up to 7\% in precision of the inferred specifications, with respect to the ground-truths gathered from the evaluation benchmarks, without affecting recall.
\end{abstract}


\section{Introduction}

Software specifications are abstract descriptions of the intended behavior of software systems \cite{Ghezzi+2002}. Software specifications are crucial to relate user needs to software behavior, and to perform different software analyses, such as program verification~\cite{Cok+2005}, test generation~\cite{DBLP:conf/icst/AbadABCFGMMRV13,DamorimETAL-ASE2006,DBLP:conf/tap/LiuMS07}, and program repair~\cite{Logozzo+2012,DBLP:journals/tse/0001FNWMZ14}. While formal specifications have many advantages and can be exploited for powerful analyses, they are known to be difficult to produce, in particular due to difficulties in formally expressing program intent. Thus, specifications are rarely seen in practice, accompanying software. 

To address this problem, researchers have developed techniques for automatically inferring specifications from other existing software elements. At the level of source code in particular, a family of approaches starting with Daikon~\cite{daikon2007}, a foundational tool in this area, generate specifications in the form of program assertions by resorting to dynamic analysis~\cite{daikon2007,gassert2020,evospex2021,specfuzzer2022}. All these approaches follow a similar approach: a mechanism is employed for generating candidate assertions for different program points (preconditions, postconditions, etc.), which are then assessed against a given test suite. Failing assertions, i.e., assertions invalidated by at least one test case, are discarded, while those assertions that are found to be valid for the provided suite are \emph{likely specifications} (or likely invariants). Likely invariants are either directly reported back to the user as likely specifications, or are the basis of what will be reported, as some tools perform \emph{a posteriori} assertion filtering to reduce redundancy~\cite{gassert2020,specfuzzer2022}. 

All these techniques face a common problem, inherent to this kind of analysis: the quality of the inferred assertions greatly depends on the quality of the test suite, and even for thorough and diverse test suites, these techniques typically report \emph{false positives}, i.e., spurious or invalid assertions which hold at the corresponding program points when observed in executions of the given test suite, but are invalid in the general case~\cite{daikon2007,evospex2021,gassert2020,specfuzzer2022}. A direct approach to deal with this issue is to inspect the produced assertions, and manually identify and discard invalid ones, before using the produced assertions for downstream analyses. However, this is time consuming and error-prone, and is exacerbated as tools for specification generation grow in specification expressiveness, and thus in the kinds of specifications they are able to report.

In this paper, we assess to what extent Large Language Models (LLMs) are effective in automating specification invalidity detection. The motivation is straightforward: if LLMs are effective at detecting invalid assertions, then the burden on the engineer can be reduced, as the likely specifications to be manually examined can be significantly reduced. To analyze the effectiveness of LLMs for this task, we take a representative tool among the above-mentioned techniques, namely SpecFuzzer~\cite{specfuzzer2022}, and introduce an additional stage in the standard generate-filter-reduce pipeline of dynamic analysis specification generation techniques: after filtering assertions via dynamic analysis, we instruct an LLM to decide whether the assertion is valid (i.e., holds at the corresponding program point for every valid run of the software under analysis) or not. Since this process is likely to be inaccurate and thus lead us to discard valid assertions, we ask the LLM to produce, when a likely invariant is deemed invalid, a test case witnessing the invalidity of the generated assertion. This approach introduces test generation \emph{after} the assertions have been produced, thus focusing test generation on the task of invalidating produced assertions already validated by dynamic analysis. The purpose is to exploit dynamic analysis on the provided test suite to reliably discard assertions, and thus focus the task of the LLM in the more challenging assertions that dynamic analysis already ``accepted''. The LLM then acts as an automated intelligent oracle, that partially simulates the traditionally human task of inspecting produced assertions, identifying false positives (spurious assertions), and removing them. By requesting invalidating test cases in our queries to the LLM, we can verify the invalidity of the assertions by executing the produced test cases and confirming assertion failure. In this way, we only remove assertions that the LLM deemed invalid and could certify such invalidity via test cases. 

We evaluate this LLM-enhanced technique on a combined benchmark from GAssert~\cite{gassert2020} and EvoSpex~\cite{evospex2021}. Our experiments consider three state-of-the-art LLMs from different families, namely, GPT-5.1 from Open AI, Llama 3.3 70B from Meta, and DeepSeek-R1. In order to more reliably assess the accuracy of the inferred specifications, we employ the curated ground truth statements provided in \cite{specfuzzer2022} for the GAssert and EvoSpex benchmarks, to implement automated ground truth checkers using SMT and SAT solvers. These checkers are used to assess accuracy in our experiments. 

Our experimental results show that the counterexamples generated by GPT-5.1 help us to discard a total of 1,877 invalid specifications inferred by SpecFuzzer, while Llama 3.3 70B discards 1,048 invalid assertions, and DeepSeek-R1 discards 2,173. This LLM-based filtering reduces in 10.09\%, 5.63\%, and 11.68\%, respectively, the noise in the inferred specifications, which otherwise would have required additional manual inspection and validation. In terms of accuracy, we observe that SpecFuzzer+GPT-5.1 obtains 74.17\% precision, 54.57\% recall, and 53.94\% F1-score, with respect to the ground-truths gathered from the benchmarks that we implemented in automated SAT/SMT checkers, a ${\sim}7\%$ precision improvement with respect to SpecFuzzer. 
Moreover, Llama 3.3 70B and DeepSeek-R1 also improve SpecFuzzer's precision by ${\sim}3.5\%$ and ${\sim}8\%$, respectively. 
These encouraging results suggest that LLMs have a significant potential in improving specification inference accuracy, thus reducing the engineering effort of the manual examination of generated specifications.

\section{Background and Related Work}

This work regards various different research areas, most notably dynamic specification inference, oracle assessment, and the use of LLMs in test generation and program analysis.

\subsection{Dynamic specification inference}

Dynamic specification inference aims at automatically producing program properties from execution scenarios. A remarkable contribution to this area, and a seminal technique and tool, is Daikon~\cite{daikon2007}. Daikon infers likely invariants by instantiating specification templates with program expressions, obtaining in this way candidate assertions, and then monitoring program executions at specific program points, such as method entrance and exit points, to assess which candidate expressions hold in all these observations at the considered program points. Daikon reports the properties that were not falsified by any test, as \emph{likely invariants} (invariants in the sense that these were found to invariantly hold at the observed program points). Daikon is very useful and has enabled various other techniques for program analysis. However, the technique is known to have expressiveness limitations, and to have its precision subject to the thoroughness of the test suite used for inference.

Various techniques have been proposed to enhance Daikon's dynamic analysis based inference, tackling in particular expressiveness limitations and the number and quality of the reported likely invariants. GAssert~\cite{gassert2020} and EvoSpex~\cite{evospex2021} employ  evolutionary approaches to generate postconditions that are validated against a given test suite. SpecFuzzer~\cite{specfuzzer2022} provides higher flexibility and expressiveness by allowing specifications to be generated via grammar-based fuzzing from a user-provided grammar, and using dynamic analysis to validate the generated specifications in the style of Daikon and other previous tools. All these tools and techniques make attempts to improve the quality of reported specifications, e.g., employing mutation-based filtering or evolution towards stronger assertions, to reduce redundancy. Despite differences in candidate generation strategies, all these techniques fundamentally rely on the strength and thoroughness of the employed test suites for dynamic analysis, to reduce the number of invalid assertions generated. All these dynamic analysis based techniques suffer from the problem of generating false positives due to the inherently partial nature of test suites as behavior specifications. 

Recently, various specification inference techniques have been proposed that directly incorporate LLMs in the specification generation process~\cite{DBLP:journals/tosem/MolinaGd25, nlp2postcondition2024, konstantinou2024llmsgeneratetestoracles, togll2024}, 
or use LLMs to generate other forms of specifications, such as 
test assertions or metamorphic relations~\cite{mrsfromreq2024, DBLP:conf/compsac/ZhangTP23}. Unlike these techniques, our approach does not modify the invariant generation or validation mechanisms themselves. Instead, it introduces a supplementary post-processing step that actively targets invalid assertions (false positives) by generating focused counterexamples, guided by an LLM, after candidate postconditions have already been inferred.

\subsection{Oracle Assessment}

The problem of assessing and improving test oracles has been widely studied in software testing research~\cite{oasis2016, oraclepolish2014, sfc2025}. OASIs~\cite{oasis2016} systematically detects these deficiencies using evolutionary search. False positives are revealed by generating test cases that violate the oracle, while false negatives are identified via mutation analysis. OASIs~\cite{oasis2016} has been used both as a quality metric and as a support tool for oracle refinement in several studies. Previous research has also focused on oracle quality, with techniques and tools such as OraclePolish~\cite{oraclepolish2014}, which employs dynamic taint analysis to detect unused inputs and brittle assertions (assertions that depend on uncontrolled inputs and thus may be invalidated). 

Our work is conceptually aligned with OASIs and OraclePolish in that we aim to generate counterexamples that expose false positives in inferred specifications. However, we differ substantially in methodology: rather than relying on evolutionary search or dynamic taint analysis, we leverage the reasoning and code generation capabilities of LLMs to directly synthesize counterexamples. Moreover, our approach is specifically tailored to dynamically inferred assertions and integrates seamlessly with existing specification inference pipelines.

\subsection{Large Language Models for Software Testing and Analysis}

Recent years have seen growing interest in leveraging LLMs for software engineering, including their application in tasks such as code generation, program repair, and test generation~\cite{LLMsurvey}. LLMs have been shown to be effective at various complex software engineering tasks, including generating unit tests, understanding program semantics, and reasoning about program behavior from source code~\cite{WangHCLWW2024}. In particular, various studies have explored the use of LLMs to generate test cases with the aim of improving code coverage or exposing bugs~\cite{foster:mutation, mhetal:fse24-llm, chattester/10.1145/3660783,chatunitest/10.1145/3663529.3663801,10.1145/3643769, WangL0J24}. Additional recent work investigates the effectiveness of LLMs for extracting software specifications from software documentation~\cite{10992315}. 

Our work differs from prior LLM-based testing and test generation methodologies in two fundamental ways. First, rather than employing LLMs for the general problem of generating test cases, we leverage them to produce tests that serve as targeted counterexamples to falsify specific inferred specifications, such as candidate program postconditions. Second, we constrain the model to output executable code that serves as a concrete witness of incorrectness, as opposed to providing unstructured natural language assessments of specification validity. This design philosophy enables the automated validation of LLM-generated responses and ensures that the resulting artifacts are executable. This implies that these responses can eventually be integrated into a dynamic loop to refine and improve an underlying specification inference process.

\section{Illustrative Example}

\begin{figure}[t]
\scriptsize
\centering
\begin{subfigure}[t]{\columnwidth}
\begin{lstlisting}[language=Java,linewidth={\linewidth},frame=tb]
public class QueueAr {
  private Object [] theArray;
  private int currentSize, front, back;

  /**
  * Get the least recently inserted item in the queue.
  * Does not alter the queue.
  */
  public Object getFront() {
    Object result;
    if(isEmpty()) {
      result = null;
    } else {
      result = theArray[front];
    }
    return result;
  }
  ...
}
\end{lstlisting}
\vspace{-1em}
\caption{Target method.}
\vspace{0.5em}
\end{subfigure}
\begin{subfigure}[t]{\columnwidth}
\begin{lstlisting}[language=Java,linewidth={\linewidth},frame=tb]
// Postcondition: (currentSize == front implies front < 1)
assert (this.currentSize != this.front || this.front < 1)
\end{lstlisting}
\vspace{-1em}
\caption{Postcondition inferred by SpecFuzzer.}
\vspace{0.5em}
\end{subfigure}
\begin{subfigure}[t]{\columnwidth}
\begin{lstlisting}[language=Java,linewidth={\linewidth},frame=tb]
@Test
public void testGetFront_1() throws Throwable {
    // Spec: (currentSize = front) implies (front < 1)
    QueueAr queue = new QueueAr(10);
    
    queue.enqueue("Item 1");
    queue.dequeue(); // This makes the queue empty again
    
    // Check the postcondition when the queue is empty
    assertTrue(queue.getFront() == null);
    
    queue.enqueue("Item 2");
    // Test that getFront() returns the correct item when the queue is not empty
    assertTrue(queue.getFront().equals("Item 2"));
}
\end{lstlisting}
\vspace{-1em}
\caption{LLM-generated counterexample.}
\vspace{0.5em}
\end{subfigure}
\caption{Situation where an LLM-generated counterexample contradicts the inferred postcondition. 
Figure a) shows the target method under analysis. Figure b) shows an 
incorrect postcondition inferred by SpecFuzzer for this method. Figure c) shows a test case generated by an LLM that contradicts the inferred postcondition, thus showing it is a false positive.}
\label{fig:example}
\end{figure}

This section presents an illustrative example to demonstrate how LLMs can enhance dynamic specification inference. Specifically, we show how LLMs can identify false positives, incorrect assertions that despite being valid in the observations considered by the dynamic analysis (restricted to the test suite being considered), are invalid in the general case. As mentioned previously, the LLMs will be instructed to provide test cases that confirm the invalidity of the identified assertions.

As a simple example, consider an array-based implementation of a queue data structure, which includes a method named \texttt{getFront} (see Figure~\ref{fig:example}(a)). The method's intended behavior is to return the least recently inserted element—the item at the logical front of the queue—while preserving the queue's state. Formally, this constitutes a postcondition: for any execution of \texttt{getFront}, if the queue is not empty, the front of the queue is returned, otherwise \texttt{null} is returned. Besides, the queue's observable state after the call must be identical to its state before the call (also part of the postcondition). The implementation logic is straightforward: it first checks if the queue is empty, in which case it returns null; otherwise, it accesses and returns the element stored at the front index of the underlying array, without modifying any structural fields.

To illustrate the automated specification inference, consider the application of SpecFuzzer~\cite{specfuzzer2022} to infer postconditions for the \texttt{getFront} method of our \texttt{QueueAr} example. Given a representative test suite, the tool generates a substantial set of 260 candidate postconditions. One such candidate, shown in Figure~\ref{fig:example}(b), asserts a relationship between internal fields: if the queue is empty (\texttt{currentSize == front}), then \texttt{front < 1}. Although this candidate postcondition may not be considered a direct characterization of the method's intended behavior, it definitely provides relevant information regarding the internal implementation of the data structure, observable at the exit point of the method under analysis. 

While the inferred postcondition is indeed informative, it is not a valid postcondition assertion for \texttt{getFront}, despite being valid for all program executions corresponding to the test suite employed for dynamic analysis. Since this is not self-evident, let us analyze the implementation of the \texttt{dequeue} method, a method that updates the \texttt{front} index, and therefore indirectly affects the possible states after executing \texttt{getFront}. As this is a circular array-based implementation of a queue,
the \texttt{dequeue} operation does not reset the \texttt{front} index to 0,
but rather increments it circularly:

\begin{lstlisting}[language=Java,linewidth={\linewidth},frame=tb]
public Object dequeue() {
  if(isEmpty()) return null;
  currentSize--;
  Object frontItem = theArray[front];
  theArray[front] = null;
  if (++front == theArray.length)
    front = 0;
  return frontItem;
}
\end{lstlisting}

\noindent
This circular increment has a direct implication: each dequeue operation advances \texttt{front} by one (modulo capacity). Consider the following execution trace, starting from an initial empty state where \texttt{currentSize = 0} and \texttt{front = 0}:

\begin{enumerate}
\item Enqueue an item: \texttt{currentSize} becomes 1; \texttt{front} remains 0.
\item Dequeue the item: \texttt{currentSize} returns to 0; \texttt{front} is incremented to 1.
\end{enumerate}
The queue is now logically empty (\texttt{currentSize = 0}), but \texttt{front = 1}. Now, perform another \texttt{enqueue}:
\begin{enumerate}
\item[3)] Enqueue a second item: \texttt{currentSize} becomes 1; \texttt{front} remains 1.
\end{enumerate}
In this final state, we have \texttt{currentSize == front == 1}. This invalidates the candidate postcondition, which states that when \texttt{currentSize == front}, \texttt{front} must be less than 1, providing a concrete counterexample.

Figure~\ref{fig:example}(c) presents a test case generated by a large language model (LLM) that encodes precisely the sequence of operations described in the previous analysis. Consequently, this test case acts as a verifiable counterexample, demonstrating the invalidity of the spurious postcondition. The generation process involved providing GPT-5.1 with three key inputs: the complete source code of the \texttt{QueueAr} class, the specific implementation of the \texttt{getFront} method, and the target candidate postcondition. We then prompted the model to first evaluate the logical validity of the postcondition and, upon determining it to be false, to synthesize an executable JUnit test case that serves as a concrete counterexample. The specific prompt structure and reasoning guidelines are detailed in Section~\ref{llm-counterexample-generation}.

The difficulty of automatically detecting spurious assertions and generating counterexamples highlights the potential of using large language models (LLMs) to improve dynamic specification inference. Our example demonstrates that LLMs can effectively identify and invalidate such flawed specifications. In the next section, we present an empirical study to systematically evaluate the effectiveness of LLMs in this concrete role.

\section{Research Questions}

Our evaluation assesses the effectiveness of LLMs in improving dynamic specification inference. We begin by investigating whether LLM-generated tests enhance the quality of inferred specifications, posing the following research question:
\begin{description}
\item[\textbf{RQ1}] Does LLM-based counterexample generation improve dynamic specification inference?
\end{description}
\noindent
To answer this question, we first prompt the LLM to judge the validity of a given postcondition for a target method; if deemed invalid, we ask the LLM to generate a test case as a counterexample. This counterexample is then integrated into the dynamic specification inference process to measure its impact on the accuracy relative to the benchmark's ground-truth specifications.

Since LLMs can make mistakes or produce tests irrespective of their relevance, accuracy improvement may be coincidental. Therefore, we shift our focus to directly evaluating the quality of the LLMs' outputs. Specifically, we assess their ability to \emph{(i)} correctly identify spurious assertions from dynamic inference output, and \emph{(ii)} generate valid counterexamples that invalidate them. This leads to the following questions:

\begin{description}

\item[\textbf{RQ2}] How effective are LLMs in identifying spurious assertions 
inferred by dynamic analysis?
  
\item[\textbf{RQ3}] How effective are LLMs in generating counterexamples
that invalidate the identified spurious assertions?
\end{description} 
\noindent
To address RQ2, we investigate the LLMs' ability to accurately identify spurious assertions that are invalid for the corresponding target method. For precise verification, we use constraint solving to determine the correctness of each inferred postcondition relative to the ground truth. For RQ3, we focus on the LLMs' capacity to generate valid counterexamples that demonstrate assertion invalidity. Here, we verify whether the generated counterexamples successfully lead to the removal of the spurious assertions during the dynamic filtering stage.

\section{Empirical Study Workflow}
\subsection{Overview}

\begin{figure*}[t]
    \centering
    \includegraphics[width=0.9\textwidth]{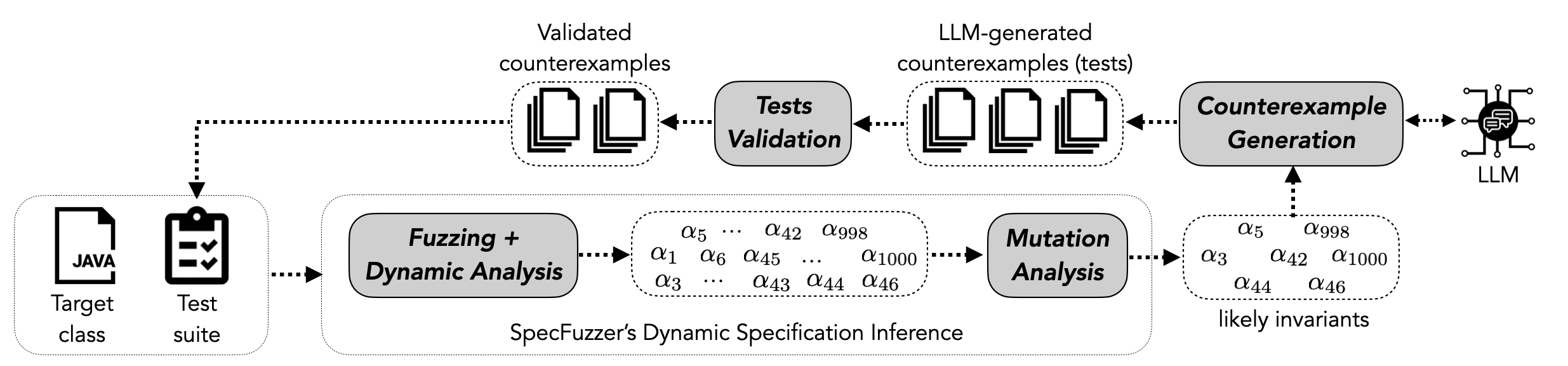}
    \caption{Overview of our empirical study workflow.}
    \label{fig:workflow}
\end{figure*}

We now provide a detailed overview of our empirical study workflow. As illustrated in Figure~\ref{fig:workflow}, our process evaluates the LLMs' potential to enhance specification inference. The workflow begins with a target Java class, aiming to infer its method postconditions as executable contract assertions. For initial inference, we employ SpecFuzzer~\cite{specfuzzer2022}, a state-of-the-art dynamic specification inference tool.

Our process continues with a target method and the set of likely postconditions dynamically inferred by SpecFuzzer. For each candidate postcondition, we query an LLM to perform a validity analysis. Specifically, we construct a prompt containing the source code of the enclosing class, the implementation of the target method, and the postcondition under scrutiny. The LLM is instructed to first reason about the postcondition's correctness. If it judges the postcondition to be invalid, the model must then generate a concrete, executable counterexample in the form of a JUnit test case that demonstrates the violation. If the postcondition is deemed valid, the process proceeds without test generation. All LLM-generated counterexample tests are subsequently aggregated and added to the original test suite. SpecFuzzer is then re-executed using this augmented suite to produce a refined—and ideally more accurate and reduced—set of inferred postconditions. The subsequent sections detail the implementation and rationale for each of these steps.

\subsection{Dynamic Specification Inference}

We use SpecFuzzer~\cite{specfuzzer2022} to infer postconditions for target Java methods. The tool operates in two phases. First, it employs grammar-based fuzzing to generate candidate postconditions and validates them against an existing test suite. Second, it performs a mutation-based filtering step to refine the candidates. During this latter phase, postconditions are grouped into clusters based on the mutants they kill. Finally, only a single representative postcondition is selected from each cluster, to form the final reported specification.

\subsection{LLM-based Counterexample Generation}
\label{llm-counterexample-generation}

\begin{figure}[t]
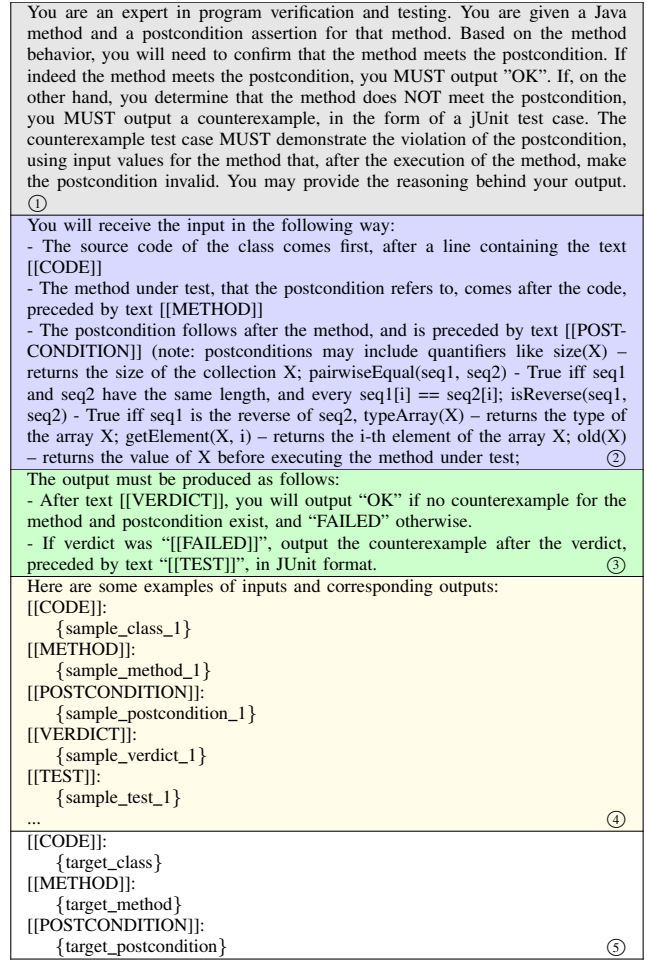

\centering
\scriptsize
\begin{tabular}{|p{0.9\columnwidth}|}
\hline
\cellcolor{gray!20}

You are an expert in program verification and testing. You are given a Java method and a postcondition assertion for that method.
Based on the method behavior, you will need to confirm that the method meets the postcondition.
If indeed the method meets the postcondition, you MUST output "OK".
If, on the other hand, you determine that the method does NOT meet the postcondition, 
you MUST output a counterexample, in the form of a jUnit test case. 
The counterexample test case MUST demonstrate the violation of the postcondition, 
using input values for the method that, after the execution of the method, 
make the postcondition invalid. You may provide the reasoning behind your output.
\hfill\circnum{1} \\
\hline
\cellcolor{blue!15}
You will receive the input in the following way: \newline
- The source code of the class comes first, after a line containing the text [[CODE]] \newline
- The method under test, that the postcondition refers to, comes after the code, preceded by text [[METHOD]] \newline
- The postcondition follows after the method, and is preceded by text [[POSTCONDITION]] (note: postconditions may include quantifiers like size(X) – returns the size of the collection X; pairwiseEqual(seq1, seq2) - True iff seq1 and seq2 have the same length, and every seq1[i] == seq2[i]; isReverse(seq1, seq2) - True iff seq1 is the reverse of seq2, typeArray(X) – returns the type of the array X; getElement(X, i) – returns the i-th element of the array X; old(X) – returns the value of X before executing the method under test; 
\hfill\circnum{2} \\
\hline
\cellcolor{green!20}
The output must be produced as follows: \newline
- After text [[VERDICT]], you will output “OK” if no counterexample 
for the method and postcondition exist, and “FAILED” otherwise. \newline
- If verdict was “[[FAILED]]”, output the counterexample 
after the verdict, preceded by text “[[TEST]]”, in JUnit format. 
\hfill\circnum{3}
\\
\hline
\cellcolor{yellow!10}
Here are some examples of inputs and corresponding outputs: \newline
[[CODE]]: \newline
\hphantom{xxx}\{sample\_class\_1\} \newline
[[METHOD]]: \newline
\hphantom{xxx}\{sample\_method\_1\} \newline
[[POSTCONDITION]]: \newline
\hphantom{xxx}\{sample\_postcondition\_1\} \newline
[[VERDICT]]: \newline
\hphantom{xxx}\{sample\_verdict\_1\} \newline
[[TEST]]: \newline
\hphantom{xxx}\{sample\_test\_1\} \newline
...
\hfill\circnum{4} \\
\hline
[[CODE]]: \newline
\hphantom{xxx}\{target\_class\} \newline
[[METHOD]]: \newline
\hphantom{xxx}\{target\_method\} \newline
[[POSTCONDITION]]: \newline
\hphantom{xxx}\{target\_postcondition\} 
\hfill\circnum{5} \\
\hline
\end{tabular}
\caption{Prompt template used for the LLM-based counterexample generation.}
\label{fig:prompt-template}
\end{figure}

To identify spurious inferred assertions and generate counterexamples that discard them, we employ state-of-the-art LLMs, with strong capabilities in code understanding and generation. To interact with the LLMs, we use a carefully designed prompt template, shown in Figure~\ref{fig:prompt-template}. The prompt aims at guiding the LLM towards determining the validity of a given postcondition
for a given method, and to generate a counterexample test 
case evidencing the violation of the postcondition, if applicable.
The prompt is composed of a system prompt and a user prompt.
The texts labeled with \circnum{1}, \circnum{2}, \circnum{3}, and \circnum{4} 
form the system prompt. 
Part \circnum{1} sets the role of the LLM as an expert in program verification and testing,
and describes the overall task of determining the validity of a postcondition
and generating counterexamples, if needed.
Parts \circnum{2} and \circnum{3} provide detailed instructions on the input format
and the expected output format, respectively.
Part \circnum{2} specifies the order in which the class code, method under test, 
and postcondition will be provided, while part \circnum{3} outlines 
how the LLM should structure its output, 
indicating that a test case should be provided only if the postcondition is violated,
in the JUnit format. 

Part \circnum{4} provides examples of inputs and corresponding outputs,
to illustrate the expected behavior of the LLM.
For space reasons, we do not show here the complete examples,
although they are available in our replication package~\cite{rep-package}.

Finally, part \circnum{5} forms the user prompt,
consisting of the actual class code, method under test, and postcondition
to be analyzed by the LLM. 

\subsection{Counterexample Validation}

The last step in our workflow validates the test cases generated by the LLM. Essentially, each generated test case is validated simply by compiling it. If the LLM-generated test does not compile, 
we instruct the LLM to fix it, with up to three attempts. If after three attempts the test case still does not compile, we discard it.
At this point, we do not execute the generated test cases, 
as they will be executed at a later stage by SpecFuzzer.

Once we have validated the generated test cases,
we incorporate all compiling test cases into the test suite 
used by SpecFuzzer. Then, we execute SpecFuzzer again
to obtain a new set of postconditions. In this paper, we only perform one SpecFuzzer re-execution, after counterexample tests have been added, we do not iterate the process. Multiple inference iterations until convergence is a research direction we plan to explore in future work.

\section{Empirical Setup}
\label{sec:setup}
\subsection{Subjects}

We consider a set of 43
Java methods that were part of the 
evaluation of 
SpecFuzzer~\cite{specfuzzer2022},
and have also been used in several related 
works on specification inference~\cite{daikon2007, gassert2020, evospex2021,seeker2023}.
This set includes methods from array-based implementations 
of stack and queue data structures,
methods from popular Java libraries 
like Apache Commons Math, Apache Commons Lang, and Google Guava, 
and JTS Topology Suite, 
and methods from classes using or implementing data structures 
including linked lists, n-ary trees, and heaps. We select these subjects because the ground truth of postconditions for these methods is already 
available, which we use as a basis to assess the correctness
of the inferred postconditions. We implemented ground truth checkers for these cases using SAT and SMT Solving. The ground truths we use are 
publicly available in our replication package~\cite{rep-package}.

\subsection{Large Language Models (LLMs)}

Our empirical evaluation considers three state-of-the-art LLMs from different families: GPT-5.1~\cite{openai-gpt5-1} from Open AI, Llama 3.3 70B~\cite{llama3-3} from Meta, and DeepSeek-R1~\cite{deepseek-r1}. 
These LLMs have shown great reasoning skills, adaptability to various tasks (Mixture-of-Experts) and good computational efficiency, making them promising candidates for our purposes.

\subsection{Evaluation Metrics}

In RQ1, we investigate whether augmenting dynamic specification inference with LLM-generated counterexamples improves the overall quality of the inferred specifications. 
To this end, we infer postcondition assertions for each 
method in our subject set using both the original SpecFuzzer tool and 
our extended SpecFuzzer implementation that incorporates LLM-based counterexample generation,
referred to as SpecFuzzer+GPT.
We compare the performance of SpecFuzzer and SpecFuzzer+GPT 
using standard performance metrics: precision, recall, and F1-score.
These metrics are computed by comparing the inferred assertions 
against the ground truth specifications.
While in previous studies these metrics were partially computed 
manually~\cite{specfuzzer2022}, in our study we fully automate this process
by formally specifying the ground truth assertions.
For subjects with numeric or boolean assertions, 
we use the Z3 SMT solver~\cite{DBLP:conf/tacas/MouraB08}, available in~\cite{z3-site}.
For more complex cases involving data structure implementations, 
we use the Alloy Analyzer~\cite{alloy-site}, 
the automated specification analysis tool
for the Alloy specification language~\cite{DBLP:journals/cacm/Jackson19}. 

Given a set $G$ of ground truth postcondition assertions for a 
subject method, we compute the precision of a set $A$ of inferred 
assertions as follows: 

\[\text{Precision} = \frac{|\{ a \in A : G \models a \}|}{|A|}\] 

\noindent
That is, precision is the ratio of inferred assertions that 
are implied by the ground truth.
To compute recall, we discard the set of invalid assertions $I \subseteq A$, the assertions in $A$ that are not implied by the ground truth $G$,
and compute recall as follows:
\[\text{Recall} = \frac{|\{ g \in G : (A \setminus I) \models g \}|}{|G|}\]

\noindent
Thus, recall is the ratio of ground truth assertions that
are implied by the valid inferred assertions (i.e., those in $A \setminus I$).
Finally, F1-score is computed as the harmonic mean of precision and recall.

In RQ2 and RQ3, we evaluate the quality and consistency of the LLM's outputs (see Figure~\ref{fig:prompt-template} for further details on the labels below). For RQ2, we measure the accuracy of the \texttt{[[VERDICT]]} provided by the LLM. Using the correct assertions identified in RQ1 as ground truth, we report precision and recall. We consider a true positive an assertion where the \texttt{[[VERDICT]]} was \texttt{FAILED} and the dynamic analysis correctly discarded it; otherwise, it is a false positive. A true negative is an assertion where the \texttt{[[VERDICT]]} was \texttt{OK} and the dynamic analysis retained it; otherwise, it is a false negative.  

In RQ3, we go one step further: for each case where the LLM's \texttt{[[VERDICT]]} was \texttt{FAILED} and it generated a \texttt{[[TEST]]}, we measure whether that counterexample indeed discards the judged invalid assertion. This allows us to assess the reasoning abilities of the LLMs and to determine whether some incorrect assertions were discarded collaterally by other tests rather than by the generated counterexamples.

\subsection{Implementation}

To run our experiments, we  implemented the workflow in Figure~\ref{fig:workflow} as a Python prototype on top of SpecFuzzer~\cite{specfuzzer2022}, adding an LLM-based counterexample generation stage after Specfuzzer's standard generate-filter-reduce pipeline. All scripts and data are in our replication package~\cite{rep-package}.

\section{Experimental Results}
\label{results}

In this section we present the results obtained when running
the GPT-5.1 LLM, and later in Section~\ref{sec:generalisation} we discuss the results with Llama 3.3 70B and DeepSeek-R1.

\subsection{RQ1: Specification Inference Improvement}

\newlength\tbspace
\setlength\tbspace{0.2cm}
\newcolumntype{R}{r<{\hspace{\tbspace}}}

\begin{table*}[t!]
\scriptsize
\caption{\label{spec-inference-improvement-per-method}Specification inference performance of SpecFuzzer and SpecFuzzer+GPT, 
our SpecFuzzer extension augmented with 
LLM-based counterexample generation using the GPT-5.1 model.}.
\centering
\setlength{\tabcolsep}{0.1em}
\renewcommand{\arraystretch}{1.2} 
\begin{tabular}{lR|rR|rR|rR|rr}
\toprule
\multirow{2}{*}{\textbf{Subject}} &
\multirow{2}{*}{\textbf{\#GT}} &
\multicolumn{2}{c}{\textbf{\#Tests}} &
\multicolumn{2}{c}{\textbf{Precision(\%)}} &
\multicolumn{2}{c}{\textbf{Recall(\%)}} &
\multicolumn{2}{c}{\textbf{F1-Score}}\\
\cmidrule(r{\tbspace}){3-4} \cmidrule(r{\tbspace}){5-6} \cmidrule(r{\tbspace}){7-8} \cmidrule(r{\tbspace}){9-10}
& & \textbf{SpecFuzzer} & \textbf{SpecFuzzer+GPT} &
\textbf{SpecFuzzer} & \textbf{SpecFuzzer+GPT} &
\textbf{SpecFuzzer} & \textbf{SpecFuzzer+GPT} &
\textbf{SpecFuzzer} & \textbf{SpecFuzzer+GPT} \\
\midrule
SimpleMethods\_addElementToSet & 1 & 313 & 315 & 100.00 & 100.00 & 100.00 & 100.00 & 100.0 & 100.0 \\
SimpleMethods\_abs & 3 & 281 & 351 & 74.70 & 96.88 & 0.00 & 0.00 & 0.0 & 0.0 \\
SimpleMethods\_getMin & 3 & 277 & 277 & 100.00 & 100.00 & 66.67 & 66.67 & 80.0 & 80.0 \\
SimpleMethods\_incrementAt & 2 & 343 & 344 & 96.49 & 96.49 & 0.00 & 0.00 & 0.0 & 0.0 \\
StackAr\_makeEmpty & 3 & 241 & 251 & 57.81 & 58.62 & 100.00 & 100.00 & 73.26 & 73.91 \\
StackAr\_topAndPop & 5 & 264 & 289 & 69.86 & 72.30 & 40.00 & 40.00 & 50.87 & 51.51 \\
StackAr\_pop & 4 & 259 & 266 & 70.89 & 71.63 & 50.00 & 50.00 & 58.64 & 58.89 \\
StackAr\_push & 4 & 264 & 280 & 89.19 & 89.19 & 50.00 & 50.00 & 64.07 & 64.07 \\
StackAr\_top & 4 & 259 & 261 & 79.66 & 79.66 & 50.00 & 50.00 & 61.44 & 61.44 \\
QueueAr\_enqueue & 5 & 241 & 294 & 80.97 & 83.30 & 80.00 & 80.00 & 80.48 & 81.62 \\
QueueAr\_getFront & 5 & 241 & 249 & 76.47 & 74.59 & 80.00 & 80.00 & 78.2 & 77.2 \\
QueueAr\_dequeue & 5 & 241 & 260 & 65.93 & 65.93 & 20.00 & 20.00 & 30.69 & 30.69 \\
QueueAr\_dequeueAll & 5 & 241 & 260 & 80.05 & 79.95 & 80.00 & 80.00 & 80.02 & 79.97 \\
QueueAr\_makeEmpty & 5 & 264 & 328 & 85.01 & 86.15 & 60.00 & 60.00 & 70.35 & 70.74 \\
ArithmeticUtils\_subAndCheck & 1 & 271 & 275 & 100.00 & 100.00 & 100.00 & 100.00 & 100.0 & 100.0 \\
FastMathNew\_floor & 1 & 286 & 348 & 40.54 & 93.75 & 0.00 & 0.00 & 0.0 & 0.0 \\
MathUtilsNew\_copySignInt & 4 & 499 & 501 & 100.00 & 100.00 & 0.00 & 0.00 & 0.0 & 0.0 \\
BooleanUtils\_toBoolean & 1 & 499 & 501 & 100.00 & 100.00 & 100.00 & 100.00 & 100.0 & 100.0 \\
BooleanUtils\_compare & 3 & 273 & 291 & 82.35 & 82.35 & 100.00 & 100.00 & 90.32 & 90.32 \\
IntMath\_mod & 1 & 328 & 360 & 1.64 & 1.67 & 0.00 & 0.00 & 0.0 & 0.0 \\
Angle\_getTurn & 4 & 499 & 501 & 100.00 & 100.00 & 25.00 & 25.00 & 40.0 & 40.0 \\
MathUtil\_clamp & 3 & 499 & 523 & 100.00 & 100.00 & 33.33 & 33.33 & 50.0 & 50.0 \\
Envelope\_maxExtent & 4 & 273 & 275 & 9.02 & 13.30 & 0.00 & 0.00 & 0.0 & 0.0 \\
Composite\_addChild & 5 & 299 & 312 & 21.39 & 22.10 & 0.00 & 0.00 & 0.0 & 0.0 \\
DLLN\_remove & 1 & 316 & 318 & 100.00 & 100.00 & 100.00 & 100.00 & 100.0 & 100.0 \\
DLLN\_insertRight & 4 & 316 & 322 & 83.87 & 82.35 & 25.00 & 25.00 & 38.52 & 38.36 \\
Map\_count & 4 & 307 & 309 & 85.19 & 85.19 & 100.00 & 100.00 & 92.0 & 92.0 \\
Map\_remove & 5 & 307 & 309 & 61.70 & 63.04 & 20.00 & 20.00 & 30.21 & 30.37 \\
Map\_extend & 4 & 307 & 329 & 63.79 & 67.27 & 25.00 & 25.00 & 35.92 & 36.45 \\
RingBuffer\_extend & 4 & 287 & 326 & 38.00 & 71.51 & 25.00 & 25.00 & 30.16 & 37.05 \\
RingBuffer\_item & 3 & 287 & 293 & 44.94 & 65.87 & 66.67 & 66.67 & 53.69 & 66.26 \\
RingBuffer\_remove & 3 & 287 & 311 & 39.22 & 59.45 & 100.00 & 100.00 & 56.34 & 74.57 \\
RingBuffer\_count & 3 & 287 & 375 & 58.28 & 91.40 & 0.00 & 0.00 & 0.0 & 0.0 \\
RingBuffer\_wipeOut & 3 & 287 & 302 & 53.25 & 63.58 & 100.00 & 100.00 & 69.5 & 77.74 \\
Polyupdate\_a1 & 2 & 271 & 489 & 1.52 & 4.46 & 50.00 & 50.00 & 2.95 & 8.2 \\
Polyupdate\_sm & 1 & 271 & 315 & 21.09 & 75.61 & 100.00 & 100.00 & 34.83 & 86.11 \\
Structure\_foo & 1 & 265 & 265 & 50.00 & 50.00 & 100.00 & 100.00 & 66.67 & 66.67 \\
Structure\_setX & 1 & 265 & 289 & 80.39 & 85.82 & 100.00 & 100.00 & 89.13 & 92.37 \\
ListComp02\_insert\_r & 3 & 259 & 261 & 94.12 & 96.77 & 100.00 & 100.00 & 96.97 & 98.36 \\
ListComp02\_insert\_s & 3 & 259 & 259 & 93.75 & 93.75 & 100.00 & 100.00 & 96.77 & 96.77 \\
MaxBag\_remove & 4 & 256 & 258 & 41.67 & 41.67 & 0.00 & 0.00 & 0.0 & 0.0 \\
MaxBag\_getMax & 3 & 256 & 282 & 100.00 & 100.00 & 66.67 & 66.67 & 80.0 & 80.0 \\
MaxBag\_add & 3 & 256 & 260 & 23.81 & 23.81 & 33.33 & 33.33 & 27.78 & 27.78 \\
\midrule
AVG & & 297.70 & 320.56 & 67.83 & 74.17 & 54.57 & 54.57 & 51.39 & 53.94 \\
\bottomrule
\end{tabular}
\end{table*}

Table~\ref{spec-inference-improvement-per-method} summarizes the accuracy of the inferred post-conditions, with respect to the ground-truth, when the LLM-generated test cases (counterexamples) are used or not in the dynamic specification inference process. 
For SpecFuzzer 
and SpecFuzzer+GPT, we report the number of ground 
truth postconditions (\#GT) for each method, 
the number of tests used by each approach, 
and the performance in terms of precision, recall, and F1-score. 

\subsubsection*{Precision}

Notably, when incorporating the LLM-generated counterexamples,
we observe
a significant improvement in precision across most subjects, 
increasing from an average of 67.83\% with SpecFuzzer 
to 74.17\% with SpecFuzzer+GPT.
More precisely, we observe that the increase in 
precision can be up to 54.52\%
for certain methods, such as \texttt{Polyupdate\_sm},
with an average increase of $\sim$7\% across all methods.
In total, the LLM-generated counterexamples allow to 
increase precision for 
48.84\% of the methods (21 out of 43 methods), 
while remaining the same for the rest.

\begin{table}[t!]
\caption{\label{tab:discarded-assertions}
Counterexample test cases generated by GPT-5.1 and their impact on the
invariants inferred by SpecFuzzer in terms 
of discarded assertions.}
\begin{center}
\setlength{\tabcolsep}{0.5em}
\scriptsize
\begin{tabular}{lr|r|rrr}
\toprule
\multirow{2}{*}{\textbf{Subject}} &
\multirow{2}{*}{\textbf{\#M}} &
\multicolumn{1}{c}{\textbf{SpecFuzzer}} & 
\multicolumn{3}{c}{\textbf{SpecFuzzer+GPT (one iter.)}} \\
& & \multicolumn{1}{c|}{\textbf{\#Invs}} & 
\textbf{\#New Tests} & \textbf{\#Invs} & \textbf{Red. $\Downarrow$}  \\
\midrule
SimpleMethods & 4 & 213 & 79 & 190 & 10.8\% \\
StackAr & 5 & 1702 & 87 & 1675 & 1.59\% \\
QueueAr & 5 & 4221 & 156 & 4062 & 3.77\% \\
ArithmeticUtils & 1 & 4 & 4 & 4 & 0\% \\
FastMath & 1 & 50 & 62 & 18 & 64\% \\
MathUtils & 1 & 18 & 2 & 17 & 5.56\% \\
BooleanUtils & 2 & 60 & 20 & 59 & 1.67\% \\
IntMath & 1 & 323 & 32 & 318 & 1.55\% \\
Angle & 1 & 4 & 2 & 3 & 25\% \\
MathUtil & 1 & 19 & 24 & 18 & 5.26\% \\
Envelope & 1 & 694 & 2 & 646 & 6.92\% \\
Composite & 1 & 7623 & 13 & 7378 & 3.21\% \\
DLLN & 2 & 162 & 8 & 160 & 1.23\% \\
Map & 3 & 141 & 26 & 137 & 2.84\% \\
RingBuffer & 5 & 2281 & 172 & 1276 & 44.06\% \\
Polyupdate & 2 & 430 & 262 & 153 & 64.42\% \\
Structure & 2 & 156 & 24 & 145 & 7.05\% \\
ListComp02 & 2 & 68 & 2 & 65 & 4.41\% \\
MaxBag & 3 & 436 & 32 & 404 & 7.34\% \\
\noalign{\vskip 0.5mm}
\hline
\noalign{\vskip 0.5mm}
TOTAL & 43 & 18,605 & 1,009 & 16,728 & 10.09\% \\
\bottomrule
\end{tabular}
\end{center}
\end{table}

This increase in precision comes 
from the quality of the counterexamples generated by the LLM,
which effectively invalidate spurious postcondition 
assertions that would otherwise be retained by SpecFuzzer.
These are test cases that, when 
incorporated into the test suite 
and executed, 
lead SpecFuzzer's dynamic 
analysis to discard a greater number of invalid assertions.
In Table~\ref{tab:discarded-assertions}, we show, for 
each subject method, the number of
new test cases generated by GPT-5.1,
the number of inferred assertions before and after
incorporating the LLM-generated tests,
and the percentage of discarded assertions. 

Notably, the 1,009 new counterexample test cases
generated by GPT-5.1 allow us to invalidate and 
eliminate a total of 1,877 invalid postcondition assertions 
inferred by SpecFuzzer. 
This represents a total reduction of 10.09\% of the inferred assertions,
which directly contributes to the observed increase in precision.
On average, GPT generated 53 new test cases per method, and 
discarded 13\% of the inferred assertions.
Some subjects experienced a significant reduction,
such as \texttt{eiffel.RingBuffer} and \texttt{cozy.Polyupdate}, 
where 44.06\% and 64.42\% of the assertions were discarded, respectively. 

\begin{figure}[t]
\scriptsize
\centering
\begin{subfigure}[t]{\columnwidth}
\begin{lstlisting}[language=Java,linewidth={\linewidth},frame=tb,mathescape]
public int abs(final int x) {
  final int i = x >>> 31;
  int result = (x^($\sim$i + 1)) + i;
  assert (result >= 0); // Inferred postcondition
  return result;
}
\end{lstlisting}
\vspace{-1em}
\caption{\texttt{SimpleMethods.abs} method and a postcondition inferred by SpecFuzzer.}
\vspace{0.5em}
\end{subfigure}
\begin{subfigure}[t]{\columnwidth}
\begin{lstlisting}[language=Java,linewidth={\linewidth},frame=tb]
@Test
public void testAbs_NegativeReturn() throws Throwable {
  examples.SimpleMethods sm = new examples.SimpleMethods();
  int x = Integer.MIN_VALUE; // -2147483648
  int result = sm.abs(x);
}
\end{lstlisting}
\vspace{-1em}
\caption{Counterexample test case generated by GPT-5.1, invalidating the inferred postcondition.}
\vspace{0.5em}
\end{subfigure}
\caption{A target method with a postcondition inferred by SpecFuzzer and a counterexample generated by GPT-5.1.}
\label{fig:abs-example}
\end{figure}

As an example of the counterexamples 
that LLMs can generate,
consider the method \texttt{SimpleMethods.abs}, 
which computes the absolute value of an integer.
Figure~\ref{fig:abs-example}(a) shows the method code
along with a postcondition inferred by SpecFuzzer,
stating that the return value \texttt{result} is always non-negative.
While this postcondition seems correct at first glance,
and holds for most integer inputs,
it fails for the specific input \texttt{Integer.MIN\_VALUE} (-2147483648) due to integer overflow: in this case, this method's implementation returns \texttt{Integer.MIN\_VALUE}. Figure~\ref{fig:abs-example}(b) shows a counterexample test case
generated by GPT-5.1 that demonstrates this violation.
When this test case is executed,
it produces a negative \texttt{result} value,
thereby invalidating the inferred postcondition. This example illustrates how LLMs can effectively reason about
subtle edge cases in program behavior,
and generate test cases that expose flaws 
in the specifications inferred by dynamic analysis 
tools like SpecFuzzer.

\subsubsection*{Recall}

Given our designed workflow, in every method we analyze, the recall will necessarily remain the same when using SpecFuzzer alone, compared to using SpecFuzzer+GPT. This is due to the fact that SpecFuzzer+GPT can only contribute additional test cases. Thus, this can only make the dynamic analysis stronger,  discarding candidate assertions that were wrongly accepted by a less thorough initial test suite. This behavior is expected, as we do not rely on the LLM to decide whether an assertion has been invalidated, but on SpecFuzzer's dynamic analysis stage, which runs each test to check assertion validity. 

Notice also that SpecFuzzer+GPT's recall is limited by SpecFuzzer's specification language and inference capabilities. Thus, any valid assertion that is not produced by SpecFuzzer, will not be captured by SpecFuzzer+GPT either. Extending SpecFuzzer's recall capabilities
would require proposing new candidate assertions, which is not supported by our designed workflow.

Finally, considering both precision and recall,
we observe an overall improvement in F1-score, 
as it positively affected by the increase in 
precision.

\begin{tcolorbox}[
  colback=gray!10,
  colframe=gray!40,
  boxrule=0.4pt,
  arc=3pt,
  left=6pt,
  right=6pt,
  top=1pt,
  bottom=1pt
]
\textbf{RQ1 Answer.}
Incorporating LLM-generated counterexamples significantly improves the precision of the inferred postconditions by effectively invalidating spurious assertions. The recall remains unchanged, confirming that no valid postconditions are discarded. The overall F1-score improves as a direct result of the increased precision.
\end{tcolorbox}

\subsection{RQ2: Invalid Postcondition Identification Effectiveness}

Table~\ref{detecting-spurious-assertions} summarizes GPT-5.1's performance in identifying invalid postconditions. We observe that GPT-5.1 correctly judged as invalid 90\% of the invalid postconditions (recall), although only 55\% of those judged invalid were indeed invalid (precision). Improving precision is a challenge to address in future LLM-based counterexample generation 
approaches. 


\begin{tcolorbox}[
  colback=gray!10,
  colframe=gray!40,
  boxrule=0.4pt,
  arc=3pt,
  left=6pt,
  right=6pt,
  top=1pt,
  bottom=1pt
]
\textbf{RQ2 Answer.}
GPT-5.1 demonstrates a very high capability to correctly identify invalid postconditions, achieving high precision and recall in its judgments. When it labels a postcondition as invalid, the prediction is highly reliable.
\end{tcolorbox}

\begin{table}[t!]
\caption{\label{detecting-spurious-assertions}
GPT-5.1 accuracy in identifying invalid postconditions.}
\begin{center}
\setlength{\tabcolsep}{10pt}
\scriptsize
\begin{tabular}{l|r|r}
\toprule
\textbf{Subject} &
\textbf{Precision} & 
\textbf{Recall} \\
\midrule
Angle\_getTurn & 100 & 100\\
BooleanUtils\_toBoolean & 100 & 100\\
Composite\_addChild & 27.27 & 100\\
Envelope\_maxExtent & 100 & 100\\
FastMathNew\_floor & 64 & 100\\
IntMath\_mod & 8.33 & 33.33\\
ListComp02\_insert\_r & 100 & 100\\
Map\_extend & 27.27 & 100\\
MathUtil\_clamp & 12.5 & 100\\
MaxBag\_getMax & 81.25 & 92.86\\
Polyupdate\_a1 & 85.39 & 95\\
Polyupdate\_sm & 91.3 & 91.3\\
QueueAr\_dequeueAll & 25 & 100\\
QueueAr\_enqueue & 62.16 & 85.19\\
QueueAr\_getFront & 80 & 100\\
QueueAr\_makeEmpty & 33.33 & 66.67\\
RingBuffer\_count & 94.92 & 96.55\\
RingBuffer\_extend & 75.76 & 100\\
RingBuffer\_item & 50 & 57.14\\
RingBuffer\_remove & 22.22 & 100\\
RingBuffer\_wipeOut & 55.56 & 62.5\\
SimpleMethods\_abs & 33.33 & 100\\
StackAr\_makeEmpty & 20 & 50\\
StackAr\_pop & 33.33 & 100\\
StackAr\_push & 25 & 100\\
StackAr\_topAndPop & 57.14 & 100\\
Structure\_setX & 23.08 & 100\\
\midrule
AVG & 55.12 & 90.02\\
\bottomrule
\end{tabular}
\end{center}
\end{table}

\subsection{RQ3: Counter-example Generation Effectiveness} 

RQ3 concerns the effectiveness of GPT-5.1 in generating counter-examples for assertions judged as invalid. We observe that the LLM judged at least one candidate postcondition as invalid in 40 out of the 43 analyzed subjects. For each of the 40 subjects, GPT-5.1 successfully generated at least one compilable counterexample, where on average, 87.2\% of the generated tests compiled correctly. In 27 out of the 40 cases, the counterexamples managed to invalidate at least one of the targeted candidate postconditions, discarding on average  47.76\% of the assertions judged invalid. 

It is worth remarking that some of the counterexamples produced by GPT-5.1 were very useful for identifying weaknesses in our ground-truth, and consequently improving it. For instance, the counterexample shown in Figure~\ref{fig:abs-example} for method  \texttt{abs}, was initially considered a false positive, since we believed that the assertion \texttt{result >= 0} was correct. However, GPT-5.1 realized that for that specific implementation, the assertion does not hold for \texttt{Integer.MIN\_VALUE} (actually, it is the only value for which it does not hold). In this case, the ground-truth was revised to match this behavior.  


\begin{tcolorbox}[
  colback=gray!10,
  colframe=gray!40,
  boxrule=0.4pt,
  arc=3pt,
  left=6pt,
  right=6pt,
  top=1pt,
  bottom=1pt
]
\textbf{RQ3 Answer.}
GPT-5.1 successfully generates compilable counterexample test cases for a large majority of the subjects. These counterexamples are effective at invalidating targeted, spurious postcondition assertions. In some cases, inaccuracies in the ground-truth are revealed.
\end{tcolorbox}

\section{Discussion}
\label{sec:discussion}

\subsection{Generalization to other LLMs}
\label{sec:generalisation}
To assess whether our findings generalize beyond the primary model under evaluation, we replicated our experiments using two additional LLMs: Llama 3.3 70B and DeepSeek-R1. 
Notice that the same requests were made for each model to mitigate any coincidental result. 
These models were selected because they belong to different model families, are open-weights, and are representative of models with state-of-the-art capabilities in problem solving and large-scale code generation. 

Table~\ref{rates_in_all_llms} summarizes the number of LLM-generated counterexamples, the number of invalid postconditions that were discarded (and the corresponding percentages), as well as the precision, recall and F1-score obtained after adding the corresponding tests to the suites used by SpecFuzzer's dynamic analysis. First, we observe that Llama 3.3 70B generated 430 compilable counterexamples for the subjects, while GPT-5.1 generated 982 compilable tests (128\% more). 
Llama's counterexamples discarded a total of 1,048 invalid postconditions, a 5.63\% reduction of the inferred specifications, in contrast to the 1,877 discarded by GPT-5.1 (79\% more than Llama). 
Llama allowed us to obtain a precision of 71.13\%, a recall of 51.47\%, and an F1-score of 50.31\%, lower than the metrics obtained with GPT-5.1. 
These observations, given that the same number and same requests were made with these models, suggest that GPT-5.1 is more cost-effective than Llama 3.3 70B, to generate counterexamples for invalid postconditions. 

Second, in the case of the DeepSeek-R1 model, it generated 3,684 compilable counterexamples (275\% more than GPT-5.1) that discarded a total of 2,173 invalid postconditions (almost 16\% more than GPT-5.1).  
DeepSeek-R1 was more effective than GPT-5.1 in generating compilable counterexamples, but it obtains a precision of 75.62\%, a recall of 51.47\%, and an F1-score of 52.04\%, a slightly lower performance than that obtained by GPT-5.1, suggesting that GPT-5.1 was the most cost-effective LLM for the task under study. 

Our results show that open-weights LLMs obtain a  performance comparable to the commercial one, being effective at identifying incorrect postconditions and generating valid counterexamples. 

\begin{table*}[th]
\scriptsize
\centering
\caption{\label{rates_in_all_llms}Comparison between different LLMs}
\resizebox{\textwidth}{!}{
\begin{tabular}{lrrrrrr}
\toprule
 \textbf{Model} & \textbf{\#Tests} & \textbf{\#Removed Assertions} & \textbf{Red. (\% $\Downarrow$)} 
 & \textbf{Precision(\%)} & \textbf{Recall(\%)} & \textbf{F1-score}\\
\midrule
GPT-5.1         & 982   & 1,877 &  10.09\% & 74.17\% & 54.57\% & 53.94\%\\
Llama 3.3 70B   & 430   & 1,048 & 5.63\% & 71.13\%  & 54.57\% &	50.31\% \\
Deepseek-R1     & 3,684 & 2,173 & 11.68\% & 75.62\% &   54.57\% &	52.04\% \\
\bottomrule
\end{tabular}
}
\end{table*}

\subsection{Towards `static' LLM-based specification inference}

Our experiments provided concrete evidence that modern LLMs have strong reasoning skills that allow them to statically determine whether a given postcondition is valid or not, for a given method under analysis. 
Then, it is rather straightforward to devise a (static) mechanism to determine the validity or invalidity of a candidate postcondition, by directly using the LLM \texttt{[[VERDICT]]} response. This may be considered an instance of the LLM-as-a-Judge strategy~\cite{ZhengLLMAsJudge}. 

To further explore this possibility, we replicated the experiments with GPT-5.1, but restricted the number of tests generated by SpecFuzzer during the dynamic analysis to just 50 tests, approximately 15\% of the number of tests used in RQ1 (see Table~\ref{tab:discarded-assertions}). This reduces SpecFuzzer's capabilities to discard invalid postconditions, so the LLM has more invalid assertions to evaluate and discard. Our experiments show that, after analyzing 15,491 assertions, GPT-5.1 managed to eliminate 5,140 invalid postconditions. 
Overall, with this simple LLM-as-a-Judge approach~\cite{ZhengLLMAsJudge}, GPT-5.1 obtained a precision of 64.21\%, a recall of 83.08, and an F1-score of 72.44\%. These results are very promising and consistent with our previous observations, encouraging researchers to explore the direct use of LLMs for candidate specification assessment in future specification inference tools.

\section{Threats to Validity}

\paragraph{External validity}

A potential threat to the validity of this study concerns the specification inference tools and LLMs used for the analysis. To mitigate this risk, we relied on SpecFuzzer, one of the state-of-the-art tools for dynamic specification inference, that has been more effective than related tools like Daikon~\cite{daikon2007}, GAssert~\cite{gassert2020} and EvoSpex~\cite{evospex2021}. In addition, we evaluated multiple LLMs from different model families to ensure that our findings are consistent and generalizable. 

\paragraph{Internal validity}

The performance of our empirical evaluation depends on the underlying tooling and LLMs. Consequently, a potential threat to internal validity arises from variations in model capability as well as from the nondeterministic nature of LLM outputs. To mitigate these threats, we evaluated multiple models from different families, reducing the risk that our findings are specific to a specific model. Additionally, to limit nondeterminism, we configured all models with a low temperature setting (0.1), aiming to make the generation process as deterministic as possible.

\paragraph{Construct validity}

A potential threat to validity is data leakage, i.e., whether the LLMs have already seen our evaluation set during training. However, the risk is low for two reasons. First, the candidate specifications are generated at runtime by SpecFuzzer using fuzzing, so the specific assertions the LLMs must assess are unlikely to appear in any training corpus. While ground truth assertions exist, we use constraint solving to check semantic consistency since the syntactic form may differ substantially from any seen example. Second, the LLMs are instructed to generate novel test cases that invalidate candidate postconditions, requiring reasoning and creating tests likely not existing for the project under analysis.

\section{Conclusions}

In this paper, we empirically evaluated whether the test generation capabilities of LLMs can improve the specification inference process. We specifically instructed three state-of-the-art LLMs, namely GPT-5.1, Llama 3.3 70B, and DeepSeek-R1, to determine whether a candidate assertion is a valid postcondition for the method under test, and to generate a counterexample in the form of a JUnit test if it is judged invalid. These counterexamples can then be used to augment the test suites employed by SpecFuzzer during dynamic analysis, helping it to discard invalid postconditions.

Our empirical evaluation on a diverse benchmark of Java methods showed that incorporating LLM-generated counterexamples leads to a substantial improvement in specification quality. In particular, GPT-5.1 led to an average increase of approximately 7 percentage points in precision, while recall remained effectively unchanged. This result indicates that GPT-5.1 can successfully eliminate invalid assertions without discarding valid properties from the ground truth. Consequently, the overall quality of inferred specifications improved, as reflected by higher F1-scores across most subjects. We also showed that this observation generalizes to other LLMs. Llama 3.3 70B discarded 5.63\% of the assertions deemed invalid, as it generated fewer compilable counterexamples than GPT-5.1. DeepSeek-R1 generated the highest number of compilable counterexamples (275\% more than GPT-5.1), discarding 11.68\% of invalid assertions (a 16\% improvement over GPT-5.1), although at a higher computational cost.

We also observed that LLMs can effectively identify invalid postconditions, achieving 90\% recall. However, precision remains at only 55\%, meaning that nearly half of the assertions flagged as invalid are actually valid. This leads to wasted resources, as the LLMs attempt to generate counterexamples for correct postconditions. Thus, while LLMs show promise in detecting actual issues, further improvements are needed to reduce false positives and make the process more efficient.



Overall, this work demonstrates that LLM-based counterexample generation is a practical and effective mechanism for improving dynamic specification inference. 

\section*{Acknowledgments}

We thank the anonymous reviewers for their valuable feedback. This work has been partially supported by Luxembourg's Ministry of  Economy through RDI Law project \emph{``Innovations for 21st Century Assessment Authoring''}, by the Luxembourg National Research Fund (FNR) PEARL program (grant agreement 16544475), by Argentina’s ANPCyT through grant 2021-4862, by China's State Administration of Foreign Experts Affairs through project \emph{``Trustworthy Evolution of LLM-generated Models''}, and by EU’s Marie Sklodowska-Curie grant No. 101008233 (MISSION).

\newpage

\bibliographystyle{plain}
\bibliography{references}

@inproceedings{ZhengLLMAsJudge,
author = {Zheng, Lianmin and Chiang, Wei-Lin and Sheng, Ying and Zhuang, Siyuan and Wu, Zhanghao and Zhuang, Yonghao and Lin, Zi and Li, Zhuohan and Li, Dacheng and Xing, Eric P. and Zhang, Hao and Gonzalez, Joseph E. and Stoica, Ion},
title = {Judging {LLM}-as-a-judge with {MT}-bench and {Chatbot Arena}},
year = {2023},
publisher = {Curran Associates Inc.},
address = {Red Hook, NY, USA},
booktitle = {Proceedings of the 37th International Conference on Neural Information Processing Systems},
location = {New Orleans, LA, USA}
}

@INPROCEEDINGS{10992315,
  author={Xie, Danning and Yoo, Byoungwoo and Jiang, Nan and Kim, Mijung and Tan, Lin and Zhang, Xiangyu and Lee, Judy S.},
  booktitle={2025 IEEE International Conference on Software Analysis, Evolution and Reengineering (SANER)}, 
  title={How Effective are Large Language Models in Generating Software Specifications?}, 
  year={2025},
  volume={},
  number={},
  pages={1-12},
  doi={10.1109/SANER64311.2025.00014}}

@inproceedings{foster:mutation,
  title={Mutation-Guided LLM-based Test Generation at Meta},
  author={Foster, Christopher and Gulati, Abhishek and Harman, Mark and Harper, Inna and Mao, Ke and Ritchey, Jillian and Robert, Herv{\'e} and Sengupta, Shubho},
  booktitle={2025 {ACM} Conference on Foundations of Software Engineering ({FSE 2025})},
  year={2025},
  organization="{ACM}",
  note={Also available as arXiv preprint arXiv:2501.12862},
}

@inproceedings{mhetal:fse24-llm,
title = "Automated unit test improvement using {Large Language Models} at {Meta}",
author  = "Nadia Alshahwan and Jubin Chheda and Anastasia Finegenova and Mark Harman and Alexandru Marginean and Shubho Sengupta and Eddy Wang",
booktitle    = "{ACM} International Conference on the Foundations of Software Engineering ({FSE 2024})",
month = "July",
year = 2024,
location = "Porto de Galinhas, Brazil, Brazil",
}

@article{chattester/10.1145/3660783,
author = {Yuan, Zhiqiang and Liu, Mingwei and Ding, Shiji and Wang, Kaixin and Chen, Yixuan and Peng, Xin and Lou, Yiling},
title = {Evaluating and Improving ChatGPT for Unit Test Generation},
year = {2024},
issue_date = {July 2024},
publisher = {Association for Computing Machinery},
address = {New York, NY, USA},
volume = {1},
number = {FSE},
url = {https://doi.org/10.1145/3660783},
doi = {10.1145/3660783},
journal = {Proc. ACM Softw. Eng.},
month = jul,
articleno = {76},
numpages = {24}
}

@inproceedings{chatunitest/10.1145/3663529.3663801,
author = {Chen, Yinghao and Hu, Zehao and Zhi, Chen and Han, Junxiao and Deng, Shuiguang and Yin, Jianwei},
title = {ChatUniTest: A Framework for LLM-Based Test Generation},
year = {2024},
isbn = {9798400706585},
publisher = {Association for Computing Machinery},
address = {New York, NY, USA},
url = {https://doi.org/10.1145/3663529.3663801},
doi = {10.1145/3663529.3663801},
booktitle = {Companion Proceedings of the 32nd ACM International Conference on the Foundations of Software Engineering},
pages = {572–576},
numpages = {5},
location = {Porto de Galinhas, Brazil},
series = {FSE 2024}
}

@inproceedings{WangL0J24,
  author       = {Zejun Wang and
                  Kaibo Liu and
                  Ge Li and
                  Zhi Jin},
  editor       = {Vladimir Filkov and
                  Baishakhi Ray and
                  Minghui Zhou},
  title        = {{HITS:} High-coverage LLM-based Unit Test Generation via Method Slicing},
  booktitle    = {Proceedings of the 39th {IEEE/ACM} International Conference on Automated
                  Software Engineering, {ASE} 2024, Sacramento, CA, USA, October 27
                  - November 1, 2024},
  pages        = {1258--1268},
  publisher    = {{ACM}},
  year         = {2024},
  url          = {https://doi.org/10.1145/3691620.3695501},
  doi          = {10.1145/3691620.3695501},
}

@article{10.1145/3643769,
author = {Ryan, Gabriel and Jain, Siddhartha and Shang, Mingyue and Wang, Shiqi and Ma, Xiaofei and Ramanathan, Murali Krishna and Ray, Baishakhi},
title = {Code-Aware Prompting: A Study of Coverage-Guided Test Generation in Regression Setting using LLM},
year = {2024},
issue_date = {July 2024},
publisher = {Association for Computing Machinery},
address = {New York, NY, USA},
volume = {1},
number = {FSE},
url = {https://doi.org/10.1145/3643769},
doi = {10.1145/3643769},
journal = {Proc. ACM Softw. Eng.},
month = jul,
articleno = {43},
numpages = {21}
}

@article{LLMsurvey,
author = {Hou, Xinyi and Zhao, Yanjie and Liu, Yue and Yang, Zhou and Wang, Kailong and Li, Li and Luo, Xiapu and Lo, David and Grundy, John and Wang, Haoyu},
title = {Large Language Models for Software Engineering: A Systematic Literature Review},
year = {2024},
publisher = {Association for Computing Machinery},
address = {New York, NY, USA},
issn = {1049-331X},
url = {https://doi.org/10.1145/3695988},
doi = {10.1145/3695988},
note = {Just Accepted},
journal = {ACM Trans. Softw. Eng. Methodol.},
month = sep
}

@article{WangHCLWW2024,
author = {Wang, Junjie and Huang, Yuchao and Chen, Chunyang and Liu, Zhe and Wang, Song and Wang, Qing},
title = {Software Testing With Large Language Models: Survey, Landscape, and Vision},
year = {2024},
issue_date = {April 2024},
publisher = {IEEE Press},
volume = {50},
number = {4},
issn = {0098-5589},
url = {https://doi.org/10.1109/TSE.2024.3368208},
doi = {10.1109/TSE.2024.3368208},
journal = {IEEE Trans. Softw. Eng.},
month = feb,
pages = {911–936},
numpages = {26}
}

@inproceedings{DamorimETAL-ASE2006,
  author    = {Marcelo d'Amorim and
               Carlos Pacheco and
               Tao Xie and
               Darko Marinov and
               Michael D. Ernst},
  title     = {An Empirical Comparison of Automated Generation and Classification
               Techniques for Object-Oriented Unit Testing},
  booktitle = {21st {IEEE/ACM} International Conference on Automated Software Engineering
               {(ASE} 2006), 18-22 September 2006, Tokyo, Japan},
  pages     = {59--68},
  publisher = {{IEEE} Computer Society},
  year      = {2006},
  url       = {https://doi.org/10.1109/ASE.2006.13},
  doi       = {10.1109/ASE.2006.13},
  bibsource = {dblp computer science bibliography, https://dblp.org}
}

@misc{togll2024,
      title={TOGLL: Correct and Strong Test Oracle Generation with LLMs}, 
      author={Soneya Binta Hossain and Matthew Dwyer},
      year={2024},
      eprint={2405.03786},
      archivePrefix={arXiv},
      primaryClass={cs.SE},
      url={https://arxiv.org/abs/2405.03786}, 
}

@inproceedings{DBLP:conf/tacas/MouraB08,
  author       = {Leonardo Mendon{\c{c}}a de Moura and
                  Nikolaj S. Bj{\o}rner},
  editor       = {C. R. Ramakrishnan and
                  Jakob Rehof},
  title        = {{Z3:} An Efficient {SMT} Solver},
  booktitle    = {Tools and Algorithms for the Construction and Analysis of Systems,
                  14th International Conference, {TACAS} 2008, Held as Part of the Joint
                  European Conferences on Theory and Practice of Software, {ETAPS} 2008,
                  Budapest, Hungary, March 29-April 6, 2008. Proceedings},
  series       = {Lecture Notes in Computer Science},
  volume       = {4963},
  pages        = {337--340},
  publisher    = {Springer},
  year         = {2008},
  url          = {https://doi.org/10.1007/978-3-540-78800-3\_24},
  doi          = {10.1007/978-3-540-78800-3\_24},
  bibsource    = {dblp computer science bibliography, https://dblp.org}
}

@article{DBLP:journals/tosem/MolinaGd25,
  author       = {Facundo Molina and
                  Alessandra Gorla and
                  Marcelo d'Amorim},
  title        = {Test Oracle Automation in the Era of LLMs},
  journal      = {{ACM} Trans. Softw. Eng. Methodol.},
  volume       = {34},
  number       = {5},
  pages        = {150:1--150:24},
  year         = {2025},
  url          = {https://doi.org/10.1145/3715107},
  doi          = {10.1145/3715107},
  bibsource    = {dblp computer science bibliography, https://dblp.org}
}

@misc{alloy-site,
      title={Alloy Analyzer}, 
      author={},
      year={2026},
      url = {https://alloytools.org/},
      howpublished = {\url{https://alloytools.org/}}
}

@misc{z3-site,
      title={Z3}, 
      author={},
      year={2026},
      url = {https://github.com/Z3Prover/z3},
      howpublished = {\url{https://github.com/Z3Prover/z3}}
}

@misc{openai-gpt5-1,
      title={GPT-5.1}, 
      author={},
      year={2026},
      url = {https://openai.com/index/gpt-5-1/},
      howpublished = {\url{https://openai.com/index/gpt-5-1/}}
}

@misc{deepseek-r1,
      title={DeepSeek-R1}, 
      author={},
      year={2026},
      url = {https://github.com/deepseek-ai/DeepSeek-R1},
      howpublished = {\url{https://github.com/deepseek-ai/DeepSeek-R1}}
}

@misc{llama3-3,
      title={Llama 3.3}, 
      author={},
      year={2026},
      url = {https://www.llama.com/docs/model-cards-and-prompt-formats/llama3_3/},
      howpublished = {\url{https://www.llama.com/docs/model-cards-and-prompt-formats/llama3_3/}}
}

@misc{rep-package,
      title={Replication package of our study}, 
      author={},
      year={2026},
      url = {https://zenodo.org/records/18899070},
      howpublished = {\url{https://zenodo.org/records/18899070}}
}

@inproceedings{Logozzo+2012,
  author    = {Francesco Logozzo and
               Thomas Ball},
  editor    = {Gary T. Leavens and
               Matthew B. Dwyer},
  title     = {Modular and verified automatic program repair},
  booktitle = {Proceedings of the 27th Annual {ACM} {SIGPLAN} Conference on Object-Oriented
               Programming, Systems, Languages, and Applications, {OOPSLA} 2012,
               part of {SPLASH} 2012, Tucson, AZ, USA, October 21-25, 2012},
  pages     = {133--146},
  publisher = {{ACM}},
  year      = {2012},
  url       = {https://doi.org/10.1145/2384616.2384626},
  doi       = {10.1145/2384616.2384626},
  bibsource = {dblp computer science bibliography, https://dblp.org}
}

@inproceedings{seeker2023,
  author       = {Aayush Garg and
                  Renzo Degiovanni and
                  Facundo Molina and
                  Maxime Cordy and
                  Nazareno Aguirre and
                  Mike Papadakis and
                  Yves Le Traon},
  title        = {Enabling Efficient Assertion Inference},
  booktitle    = {34th {IEEE} International Symposium on Software Reliability Engineering,
                  {ISSRE} 2023, Florence, Italy, October 9-12, 2023},
  pages        = {623--634},
  publisher    = {{IEEE}},
  year         = {2023},
  url          = {https://doi.org/10.1109/ISSRE59848.2023.00039},
  doi          = {10.1109/ISSRE59848.2023.00039},
  bibsource    = {dblp computer science bibliography, https://dblp.org}
}

@article{DBLP:journals/cacm/Jackson19,
  author       = {Daniel Jackson},
  title        = {Alloy: a language and tool for exploring software designs},
  journal      = {Commun. {ACM}},
  volume       = {62},
  number       = {9},
  pages        = {66--76},
  year         = {2019},
  url          = {https://doi.org/10.1145/3338843},
  doi          = {10.1145/3338843},
  bibsource    = {dblp computer science bibliography, https://dblp.org}
}

@article{daikon2007,
  author    = {Michael D. Ernst and
               Jeff H. Perkins and
               Philip J. Guo and
               Stephen McCamant and
               Carlos Pacheco and
               Matthew S. Tschantz and
               Chen Xiao},
  title     = {The Daikon system for dynamic detection of likely invariants},
  journal   = {Sci. Comput. Program.},
  volume    = {69},
  number    = {1-3},
  pages     = {35--45},
  year      = {2007},
  url       = {https://doi.org/10.1016/j.scico.2007.01.015},
  doi       = {10.1016/j.scico.2007.01.015},
  bibsource = {dblp computer science bibliography, https://dblp.org}
}

@inproceedings{evospex2021,
  author    = {Facundo Molina and
               Pablo Ponzio and
               Nazareno Aguirre and
               Marcelo F. Frias},
  title     = {EvoSpex: An Evolutionary Algorithm for Learning Postconditions},
  booktitle = {43rd {IEEE/ACM} International Conference on Software Engineering,
               {ICSE} 2021, Madrid, Spain, 22-30 May 2021},
  pages     = {1223--1235},
  publisher = {{IEEE}},
  year      = {2021},
  url       = {https://doi.org/10.1109/ICSE43902.2021.00112},
  doi       = {10.1109/ICSE43902.2021.00112},
  bibsource = {dblp computer science bibliography, https://dblp.org}
}

@inproceedings{specfuzzer2022,
  author       = {Facundo Molina and
                  Marcelo d'Amorim and
                  Nazareno Aguirre},
  title        = {Fuzzing Class Specifications},
  booktitle    = {44th {IEEE/ACM} 44th International Conference on Software Engineering,
                  {ICSE} 2022, Pittsburgh, PA, USA, May 25-27, 2022},
  pages        = {1008--1020},
  publisher    = {{ACM}},
  year         = {2022},
  url          = {https://doi.org/10.1145/3510003.3510120},
  doi          = {10.1145/3510003.3510120},
  bibsource    = {dblp computer science bibliography, https://dblp.org}
}

@book{Ghezzi+2002,
     author = {Ghezzi, Carlo and Jazayeri, Mehdi and Mandrioli, Dino},
      title = {Fundamentals of Software Engineering},
       year = {2002},
        isbn = {0133056996},
         edition = {2nd},
          publisher = {Prentice Hall PTR},
           address = {Upper Saddle River, NJ, USA},
}

@inproceedings{sfc2025,
  author       = {Facundo Molina and
                  Nazareno Aguirre and
                  Alessandra Gorla},
  title        = {State Field Coverage: {A} Metric for Oracle Quality},
  booktitle    = {40th {IEEE/ACM} International Conference on Automated Software Engineering,
                  {ASE} 2025, Seoul, Korea, Republic of, November 16-20, 2025},
  pages        = {2707--2719},
  publisher    = {{IEEE}},
  year         = {2025},
  url          = {https://doi.org/10.1109/ASE63991.2025.00222},
  doi          = {10.1109/ASE63991.2025.00222},
  bibsource    = {dblp computer science bibliography, https://dblp.org}
}

@inproceedings{oasis2016,
  author    = {Gunel Jahangirova and
               David Clark and
               Mark Harman and
               Paolo Tonella},
  editor    = {Andreas Zeller and
               Abhik Roychoudhury},
  title     = {Test oracle assessment and improvement},
  booktitle = {Proceedings of the 25th International Symposium on Software Testing
               and Analysis, {ISSTA} 2016, Saarbr{\"{u}}cken, Germany, July
               18-20, 2016},
  pages     = {247--258},
  publisher = {{ACM}},
  year      = {2016},
  url       = {https://doi.org/10.1145/2931037.2931062},
  doi       = {10.1145/2931037.2931062},
  bibsource = {dblp computer science bibliography, https://dblp.org}
}

@inproceedings{DBLP:conf/icst/AbadABCFGMMRV13,
  author    = {Pablo Abad and
               Nazareno Aguirre and
               Valeria S. Bengolea and
               Daniel Alfredo Ciolek and
               Marcelo F. Frias and
               Juan P. Galeotti and
               Tom Maibaum and
               Mariano M. Moscato and
               Nicol{\'{a}}s Rosner and
               Ignacio Vissani},
  title     = {Improving Test Generation under Rich Contracts by Tight Bounds and
               Incremental {SAT} Solving},
  booktitle = {Sixth {IEEE} International Conference on Software Testing, Verification
               and Validation, {ICST} 2013, Luxembourg, Luxembourg, March 18-22,
               2013},
  pages     = {21--30},
  publisher = {{IEEE} Computer Society},
  year      = {2013},
  url       = {https://doi.org/10.1109/ICST.2013.46},
  doi       = {10.1109/ICST.2013.46},
  bibsource = {dblp computer science bibliography, https://dblp.org}
}

@article{Cok+2005,
  author    = {Gary T. Leavens and
               Yoonsik Cheon and
               Curtis Clifton and
               Clyde Ruby and
               David R. Cok},
  title     = {How the design of {JML} accommodates both runtime assertion checking
               and formal verification},
  journal   = {Sci. Comput. Program.},
  volume    = {55},
  number    = {1-3},
  pages     = {185--208},
  year      = {2005},
  url       = {https://doi.org/10.1016/j.scico.2004.05.015},
  doi       = {10.1016/j.scico.2004.05.015},
  bibsource = {dblp computer science bibliography, https://dblp.org}
}

@inproceedings{oraclepolish2014,
  author       = {Chen Huo and
                  James Clause},
  editor       = {Shing{-}Chi Cheung and
                  Alessandro Orso and
                  Margaret{-}Anne D. Storey},
  title        = {Improving oracle quality by detecting brittle assertions and unused
                  inputs in tests},
  booktitle    = {Proceedings of the 22nd {ACM} {SIGSOFT} International Symposium on
                  Foundations of Software Engineering, (FSE-22), Hong Kong, China, November
                  16 - 22, 2014},
  pages        = {621--631},
  publisher    = {{ACM}},
  year         = {2014},
  url          = {https://doi.org/10.1145/2635868.2635917},
  doi          = {10.1145/2635868.2635917},
  bibsource    = {dblp computer science bibliography, https://dblp.org}
}

@inproceedings{gassert2020,
author = {Terragni, Valerio and Jahangirova, Gunel and Tonella, Paolo and Pezz\`{e}, Mauro},
title = {Evolutionary Improvement of Assertion Oracles},
year = {2020},
isbn = {9781450370431},
publisher = {Association for Computing Machinery},
address = {New York, NY, USA},
url = {https://doi.org/10.1145/3368089.3409758},
doi = {10.1145/3368089.3409758},
booktitle = {Proceedings of the 28th ACM Joint Meeting on European Software Engineering Conference and Symposium on the Foundations of Software Engineering},
pages = {1178–1189},
numpages = {12},
location = {Virtual Event, USA},
series = {ESEC/FSE 2020}
}

@article{DBLP:journals/tse/0001FNWMZ14,
  author    = {Yu Pei and
               Carlo A. Furia and
               Martin Nordio and
               Yi Wei and
               Bertrand Meyer and
               Andreas Zeller},
  title     = {Automated Fixing of Programs with Contracts},
  journal   = {{IEEE} Trans. Software Eng.},
  volume    = {40},
  number    = {5},
  pages     = {427--449},
  year      = {2014},
  url       = {https://doi.org/10.1109/TSE.2014.2312918},
  doi       = {10.1109/TSE.2014.2312918},
  bibsource = {dblp computer science bibliography, https://dblp.org}
}

@article{nlp2postcondition2024,
  author       = {Madeline Endres and
                  Sarah Fakhoury and
                  Saikat Chakraborty and
                  Shuvendu K. Lahiri},
  title        = {Can Large Language Models Transform Natural Language Intent into Formal
                  Method Postconditions?},
  journal      = {Proc. {ACM} Softw. Eng.},
  volume       = {1},
  number       = {{FSE}},
  pages        = {1889--1912},
  year         = {2024},
  url          = {https://doi.org/10.1145/3660791},
  doi          = {10.1145/3660791},
  bibsource    = {dblp computer science bibliography, https://dblp.org}
}

@misc{konstantinou2024llmsgeneratetestoracles,
      title={Do LLMs generate test oracles that capture the actual or the expected program behaviour?}, 
      author={Michael Konstantinou and Renzo Degiovanni and Mike Papadakis},
      year={2024},
      eprint={2410.21136},
      archivePrefix={arXiv},
      primaryClass={cs.SE},
      url={https://arxiv.org/abs/2410.21136}, 
}

@inproceedings{mrsfromreq2024,
author={Shin, Seung Yeob
and Pastore, Fabrizio
and Bianculli, Domenico
and Baicoianu, Alexandra},
editor={Bertolino, Antonia
and Pascoal Faria, Jo{\~a}o
and Lago, Patricia
and Semini, Laura},
title={Towards Generating Executable Metamorphic Relations Using Large Language Models},
booktitle={Quality of Information and Communications Technology},
year={2024},
publisher={Springer Nature Switzerland},
pages={126--141},
}

@inproceedings{DBLP:conf/compsac/ZhangTP23,
  author       = {Yifan Zhang and
                  Dave Towey and
                  Matthew Pike},
  editor       = {Hossain Shahriar and
                  Yuuichi Teranishi and
                  Alfredo Cuzzocrea and
                  Moushumi Sharmin and
                  Dave Towey and
                  A. K. M. Jahangir Alam Majumder and
                  Hiroki Kashiwazaki and
                  Ji{-}Jiang Yang and
                  Michiharu Takemoto and
                  Nazmus Sakib and
                  Ryohei Banno and
                  Sheikh Iqbal Ahamed},
  title        = {Automated Metamorphic-Relation Generation with ChatGPT: An Experience
                  Report},
  booktitle    = {47th {IEEE} Annual Computers, Software, and Applications Conference,
                  {COMPSAC} 2023, Torino, Italy, June 26-30, 2023},
  pages        = {1780--1785},
  publisher    = {{IEEE}},
  year         = {2023},
  url          = {https://doi.org/10.1109/COMPSAC57700.2023.00275},
  doi          = {10.1109/COMPSAC57700.2023.00275},
  bibsource    = {dblp computer science bibliography, https://dblp.org}
}

@inproceedings{DBLP:conf/tap/LiuMS07,
  author    = {Lisa (Ling) Liu and
               Bertrand Meyer and
               Bernd Schoeller},
  editor    = {Yuri Gurevich and
               Bertrand Meyer},
  title     = {Using Contracts and Boolean Queries to Improve the Quality of Automatic
               Test Generation},
  booktitle = {Tests and Proofs - 1st International Conference, {TAP} 2007, Zurich,
               Switzerland, February 12-13, 2007. Revised Papers},
  series    = {Lecture Notes in Computer Science},
  volume    = {4454},
  pages     = {114--130},
  publisher = {Springer},
  year      = {2007},
  url       = {https://doi.org/10.1007/978-3-540-73770-4\_7},
  doi       = {10.1007/978-3-540-73770-4\_7},
  bibsource = {dblp computer science bibliography, https://dblp.org}
}

\end{document}